\documentstyle[12pt]{article}
\def\be{\begin{equation}}
\def\ee{\end{equation}}
\def\pa{\partial}
\def\eps{\epsilon}
\begin{document}
\title  { Dual composition of odd-dimensional models}
\author {
 {Rabin Banerjee$^1$ {\footnote {rabin@bose.res.in}}} and {Sarmishtha Kumar(Chaudhuri)$^2$
 {\footnote {chaudhurisarmistha@gmail.com}}}\\
$^1$ S.N.Bose National Centre For Basic Sciences, \\SaltLake City,Block JD,
Sector III, Calcutta 700 098, India.
\\$^2$ Camellia Institute of Technology, Madhyamgram, Kolkata-700129,India}
\maketitle

\begin {abstract}
A general way of interpreting odd dimensional models as a doublet of chiral models is discussed.
Based on the equations of motion this dual composition is illustrated. Examples from quantum
 mechanics, field theory and gravity are considered. Specially the recently advocated
topologically massive gravity is analysed in some details.

\end{abstract}

\section*{I.Introduction}

Equivalent descriptions of the same physical theory are useful and play a significant role in
expanding our understanding. Aspects of a theory that are hidden in one formulation become transparent
in some other formulation. The time honored bosonisation technique in $(1+1)$ dimensions is a classic example
in this context \cite{CO}. More recent examples are provided by the AdS/CFT correspondence \cite{JM}
and the duality discussed in \cite{SW} that paved the path for a rigorous treatment of confinement
in a four dimensional theory. Apart from this direct model to model equivalence, there is another sort
of dual description where a particular theory is interpreted as a combination or a doublet of theories.
Such a description is sometimes not just desirable, but essential, in abstracting the spectrum
of the composite theory. A typical illustration is the Proca model in $(2+1)$ dimensions. The two
massive modes of this model are known to be obtained from a doublet of self-dual models
with helicity $\pm1$ \cite{BW,BK1}, a fact that was briefly suggested in \cite{DJT}.
Very recently, similar notions and concepts have been
exploited to study a new version of topologically massive gravity \cite{DM,DM1}. 
The present paper is devoted to this aspect, albeit from a more general perspective. 

Broadly speaking there are two approaches to visualise this doublet structure of a composite
theory--one based on the lagrangian formulation while the other involves the hamiltonian
formulation. The standard viewpoint in the first approach is to solder the distinct lagrangians
through a contact term \cite{BW,BK1} while in the latter(hamiltonian) case, a canonical transformation
is found that diagonalises the hamiltonian into independent pieces \cite{DJTr,BG,BK2}. We shall
here concentrate on the lagrangian version. However we avoid the usual soldering formalism
which sometimes becomes technically involved requiring arcane field redefinitions. We adopt
a method based on equations of motion \cite{D} necessitating very simple field redefinitions that are generic
to a wide variety of models.

The basic ideas are introduced in section II where we analyse a quantum mechanical model.
This is discussed in some details because it may be interpreted as field theory in $(0+1)$ dimension
which is a precursor to field theory in $(2+1)$ dimensions. We show that the motion of a charged particle
in the presence of electric and magnetic fields is simulated by a doublet of chiral oscillators,
one moving in the clockwise while the other in the anticlockwise derection. This discussion is 
extended in section III to the case of spin $1$ vector models in $(2+1)$ dimensions.  
Source terms are also included. Then in sections IV and V we analyse exhaustively spin $2$ 
tensor models which appear in discussions \cite{DJT,BH} of linearised gravity in $(2+1)$ dimensions. 
Taking a doublet of selfdual massive spin $2$ models
considered earlier in \cite{AK}, we show that the effective theory is a new type of generalised
self dual model that has a Fierz-Pauli term, a first order Chern-Simons term and the Einstein-Hilbert
term. Subject to a specific condition, it reduces to the model taken in \cite{ DM}. Our conclusions
and final remarks are relegated to section VI.

\newpage

\section*{II.Quantum mechanical model}

We introduce our formalism by considering the two dimensional topological quantum mechanical
model governed by the Lagrangian \cite{BK1,DJTr} 
\begin{equation}
{\cal {L}} = \frac{m}{2}{\dot{x_i}}^2 + e\dot{x_i}{A_i}(x) - eV(x)\quad\quad\quad(i=1,2)
\label{lag}
\end{equation}
exhibiting the motion of a charged particle in external electric and magnetic fields. For a rotaionally
symmetric motion in a constant magnetic field and a harmonic potential,
$$
A_i = -\frac{1}{2}\epsilon_{ij}{x_j}B \quad\quad,\quad\quad V  =\frac{k}{2}{x_i}^2  \nonumber
$$
the lagrangian (\ref{lag}) simplifies to (setting e=1),

\begin{equation}
 {\cal {L}} = \frac{m}{2}{\dot{x_i}}^2 + \frac{B}{2}\epsilon_{ij}x_i
\dot{x_j} - \frac{k}{2}{x_i}^2
\label{lag1}
\end{equation} 
This Lagrangian may be interpreted as a combination of two chiral oscillators. It was shown
in \cite{BK1} where the discussion was performed at the level of the action. Here we show
how similar conclusions may be obtained in a simpler way by looking at equations of motion.
This approach will be exploited to treat the more complex examples of field theory and gravity.
\\
 Let us consider a pair of lagrangians,
\begin{eqnarray}
{\cal {L}_+}& =&  \frac{1}{2\omega_+}\epsilon_{ij}x_i\dot{x_j} - \frac{1}{2}{x_i}^2     
\label{1a} \\
{\cal{L}_-} &= & - \frac{1}{2\omega_-}\epsilon_{ij}y_i\dot{y_j}-\frac{1}{2}{y_i}^2
\label{1b}
\end{eqnarray}
where independent set of coordinates $x_i$ and $y_i$ have been used. The equations of motion
are given by,
\begin {equation}
\frac{1}{{\omega}_+}\epsilon_{ik}\dot{x_k} - x_i= 0
\label{2a}
\end{equation}
\begin{equation}
-\frac{1}{{\omega}_-}\epsilon_{ik}\dot{y_k} - y_i= 0
\label{2b}
\end{equation}
These may be put in the form,

$$\ddot{x_i} = -{{\omega_+}^2}{x_i}\quad\quad,\quad\quad \ddot{y_i} = -{{\omega_-}^2}{y_i} \nonumber$$
which are just the standard oscillator equations. Consequently (\ref{1a}) and (\ref{1b}) represent
two chiral oscillators (with frequencies $\omega_+ $ and $\omega_-$) having opposite chirality 
which is manifested by the different signs of the first term in $\cal{L}_\pm$.\\
Let us introduce a new set of variables $(f_i, g_i)$ by combining chiral ones as,
\be
 x_i+y_i = f_i, \,\,\,\,\, x_i-y_i = g_i
\label{2c}
\ee
Subtracting (\ref{2b}) from (\ref{2a}) and using the definition of $g_i$ we obtain,
\begin{equation}
\frac{1}{\omega_+}\epsilon_{ik}\dot{g_k} + \Omega\epsilon_{ik}\dot{y_k} - g_i= 0 \quad;
\quad\quad\quad\quad \Omega=\frac{1}{\omega_+} +\frac{1}{\omega_-}
\label{3a}
\end{equation}
Contracting (\ref{3a}) by ${\frac{1}{\omega_-}}\epsilon_{ip}$ yields,
\be
\frac{1}{{\omega_+}{\omega_-}}\ddot{g_p}+\frac{1}{\omega_-}\epsilon_{pi}\dot{g_i}+
\frac{\Omega}{\omega_-}\ddot{y_p}=0
\label{3b}
\ee
Taking the difference of (\ref{3a}) and (\ref{3b}) and exploiting the oscillator equation of motion 
for $y$ yields, 
\begin{equation}
\ddot{g_i} + ({\omega_+}-{\omega_-})\epsilon_{ik}\dot{g_k}+{\omega_+}{\omega_-}{g_i}=0
\label{4a}
\end{equation}
Adopting a similar analysis we can show that the other variable $f_i$ also satisfies
an identical equation,
\begin{equation}
\ddot{f_i} + ({\omega_+}-{\omega_-})\epsilon_{ik}\dot{f_k}+{\omega_+}{\omega_-}{f_i}=0
\label{4b}
\end{equation}
These equations of motion factorise in terms of their dual(chiral) components as \cite{BK1},
\be
(\epsilon_{ji}\partial_t + \omega_-\delta_{ji})(\epsilon_{ik}\partial_t - \omega_+\delta_{ik})X_k = 0 \quad\quad;
\quad\quad \quad{X_k}=({f_k},{g_k})
\label{4b.1}
\ee
Observe that the above equations of motion can be obtained from the lagrangian,
\be
{\cal L} = \frac{1}{2} {\dot{X_i}}^2 - \frac{1}{2}(\omega_+ - \omega_-)\epsilon
_{ij}X_i\dot{X_j} - \frac{1}{2}{\omega_+}{\omega_-}{X_i}^2
\label{4c}
\ee

It is now straightforward to identify this lagrangian with (\ref{lag1}) for a unit mass(m=1) by taking,
${\omega_-} - {\omega_+}=B$  and  $ {\omega_+}{\omega_-}=k $.

This shows how two chiral oscillators with distinct frequencies moving in clockwise and anticlockwise
directions simulate the motion of a charged particle in the presence of electric and magnetic fields.
Moreover the magnetic part is a consequence of the different angular frequencies. For 
${\omega_+}={\omega_-}$, this term just drops out and only the effect of the electric field is retained.
Finally, note that the obtention of (\ref{4c}) is effected by the change of variables (\ref{2c}).
Such a change will also play a crucial role in both field theory and gravity to be discussed in the subsequent
sections.


\section*{III. Vector model with different coupling}

In this section we shall extend the idea developed in the previous section to (2+1) dimensional
field theory. This is a natural extension since the earlier quantum mechanical model
may be interpreted as field theory in (0+1)dimensions. In that case the lagrangians in (\ref{1a}) 
and  (\ref{1b}) would be interpreted as the analogues of self and anti-selfdual models \cite{TPN,DJ}
in (2+1)D in the limit where all spatial derivatives are ignored.\\
Let us consider therefore the following doublet of models,
\be
{\cal L}_{SD} = {\cal L} _{+} = \frac{1}{2} f_{\mu} f^{\mu} - \frac{1}{2m_1}
\eps_{\mu \nu\lambda}f^{\mu}\pa ^{\nu} f ^\lambda + \frac{1}{2}{f_\mu}{J^\mu}
\label{2.1}
\ee 

\be
{\cal L}_{ASD} = {\cal L} _{-} = \frac{1}{2} g_{\mu} g^{\mu} + \frac{1}{2m_2} 
\eps_{\mu \nu\lambda} g^{\mu} \pa ^{\nu} g ^{\lambda} + \frac{1}{2}{g_\mu}{J^\mu}
\label{2.2} 
\ee
The different signs in the Chern-Simons piece show that they may be regarded as a
set of selfdual and anti-selfdual models \cite{TPN,DJ,BW} where we have also included source terms. The equations
of motion are given by,
\be
f_{\mu}  - \frac{1}{m_1}\eps_{\mu \nu\lambda}\pa ^{\nu}f ^{\lambda} + \frac{1}{2}J^{\mu} = 0
\label{2.3}
\ee
and
\be
g_{\mu}  + \frac{1}{m_2}\eps_{\mu \nu\lambda}\pa ^{\nu}g^{\lambda} + \frac{1}{2}J^{\mu} = 0
\label{2.4}
\ee

Proceeding as before we introduce a new set of fields which are the analogues of (\ref{2c}),
\be
 f^{\mu} +g^{\mu} = k^\mu ;\quad\quad\quad f^{\mu} -g^{\mu} = h^\mu 
\label{2.4a} 
\ee
Now adding (\ref{2.3}) and (\ref{2.4}) and substituting the old fields by the new one $k_\mu$(say)
leads to
\be
k_{\mu}  - \frac{1}{m_1}\eps_{\mu \nu\lambda}\pa ^{\nu}k ^{\lambda} +
 M\eps_{\mu \nu\lambda}\pa ^{\nu}g ^{\lambda} + J^{\mu} = 0
\label{2.5}
\ee 
where $M=(\frac{1}{m_1} + \frac{1}{m_2}) $.

Multiplying (\ref{2.5}) by $\frac{1}{m_2}\eps^{\sigma \rho\mu}\pa_{\rho}$ yields,
\be
\frac{1}{m_2}\eps^{\sigma \rho\mu}\pa_{\rho}{k_\mu} - \frac{1}{{m_1}{m_2}}\pa_{\rho}{k^{\rho\sigma}}
 + \frac{M}{m_2}\pa_{\rho}g^{\rho\sigma} + \frac{1}{m_2}\eps^{\sigma \rho\mu}\pa_{\rho}{J_\mu} = 0
\label{2.6}
\ee
where, $k_{\mu \nu}= \partial_{[\mu}k_{\nu]}$ and  $g_{\mu \nu}= \partial_{[\mu}g_{\nu]}$ are antisymmetric
combinations that may be interpreted as field tensors associated with the respective fields $k_\mu $
 and $g_\mu $.

In order to eliminate the $g_\mu$ variable from (\ref{2.6}), we add (\ref{2.5}) to it and then exploit (\ref{2.4}).
One finally obtains

\be
\pa_{\rho}k^{\rho\sigma} +  ({m_2} - {m_1}) \eps^{\sigma \rho\lambda}\pa_{\rho}{k_\lambda}  - {m_2}{m_1}{k^\sigma} 
=  {m_2}{m_1}{J^\sigma} +\\
\frac{1}{2}({m_1} - {m_2})\eps^{\sigma \rho\lambda}\pa_{\rho}{J_\lambda}
\label{2.7}
\ee
The expression here is purely in terms of the new variable $k_\mu$.

Following similar steps an equation involving only the $h_\mu$ variable is obtained,
\be
\pa_{\rho}h^{\rho\sigma} +  ({m_2} - {m_1}) \eps^{\sigma \rho\lambda}\pa_{\rho}{h_\lambda}  - {m_2}{m_1}{h^\sigma} 
=  \frac{1}{2}({m_1} + {m_2})\eps^{\sigma\rho\lambda}\pa_{\rho}{J_\lambda}
\label{2.7a}
\ee
In the absence of sources the above equations display a factorisation property analogous to (\ref{4b.1}),
\begin{equation}
[g^{\rho\mu} + \frac{1}{m_2}\epsilon^{\rho\sigma\mu}\partial_{\sigma} ]
[g_{\mu\lambda} - \frac{1}{m_1}\epsilon_{\mu\nu\lambda}\partial ^{\nu} ] f^{\lambda} =0  \quad\quad;
\quad\quad\quad(f^\lambda = k^\lambda,h^\lambda)
\label{4b.2}
\end{equation}

The lagrangians which lead to (\ref{2.7}) and (\ref{2.7a}) are given by,
\begin{eqnarray}
{{\cal L}_k} = - \frac {1}{4}k_{\mu \nu}k^{\mu \nu}  - \frac {1}{2}({m_2} - {m_1})
 \eps_{\mu \nu \sigma}\partial ^{\mu}{ k^\nu}k^{\sigma} + \frac{1}{2}{m_1}{m_2}{ k_\mu}{ k^\mu}
 - {m_1}{m_2}{ k_\mu}{ J^\mu} \cr
 - \frac{1}{2}({m_2} - {m_1})\eps_{\mu \nu \sigma}{k^\mu}\partial ^{\nu}J^{\sigma}
\label{2.8}
\end{eqnarray}
and
\be
{{\cal L}_h}  = - \frac {1}{4}{ h_{\mu \nu}}{h^{\mu \nu}} - \frac {1}{2}({m_2} - {m_1})
 \eps_ {\mu \nu \sigma}\partial ^{\mu} h^{\nu}{h^\sigma}+ \frac{1}{2}{m_1}{m_2}{h_\mu}{h^\mu}
 - \frac{1}{2}({m_2}+{m_1})\eps_{\mu \nu \sigma}h^{\mu}\partial ^{\nu}J^{\sigma}
\label{2.9}
\ee

It is notable that in the absence of the sources both the effective lagrangians in
(\ref{2.8}) and (\ref{2.9}) are identical.
The first term represents the ordinary Maxwell term, the second one involving epsilon 
specifies the Chern-Simons term and the last one is a mass term.
Therefore the effective lagrangian density ${{\cal L}_h} $
or ${\cal L}_k $ gets identified with the Maxwell-Chern Simons-Proca(MCSP) model. 
This result was obtained earlier using various approaches ranging from the soldering of actions
\cite{BW,BK1} to path integral methods \cite{BK2} based on master actions. Within the hamiltonian
formalism this was achieved by using the canonical transformations \cite{BG,BK2}. Compared to these,
the present analysis is very economical and follows as a natural extension of the quantum mechanical
analysis presented earlier. Furthermore, although both $ k_\mu $ and  $ h_\mu $ yield the same free
theory, their roles are quite distinct in the presence of interactions as may be evidenced
from the different source contributions appearing in (\ref{2.8}) and (\ref{2.9}) repectively.

The dual composition is succinctly expressed by the following maps:
\be
{\cal L}_{2SD}(f,g) \Longleftrightarrow {\cal L}_{MCSP}(f \pm g)
\label{2.10}
\ee
where the left hand side indicates a doublet comprising self and anti-self dual models (\ref{2.1},\ref{2.2})
while the right hand side depicts the composite model that is expressed either in terms of the $(f+g)$
variable (\ref{2.8}) or the $(f-g)$ variable (\ref{2.9})

Let us now look into the case when ${m_1}={m_2}={m}$ . Then the epsilon term vanishes reducing the expression
for ${\cal L}_k $ or ${\cal L}_h $ to the usual Proca model. Also the source term gets considerably
simplified.

\section*{IV. Extrapolation to Gravity }  

   Here we shall implement the notions developed in the previous sections to discuss the example of
rank two tensor models which arise in various formulations of gravity. In section III we
illustrated the combination of a doublet of self-dual models with distinct masses and spins $\pm1$
to yield an effective Maxwell-Chern-Simons-Proca model. Here we consider the combination of a doublet
of spin$\pm2$ models that arise in linearised gravity.

Let us start with the action of first order selfdual massive spin 2 model as suggested in \cite{AK}

\be
S = \int{d^3}x\lbrack \frac{m}{2}\eps^{\mu\nu\lambda}{f_\mu}\,^{\alpha}\pa_{\nu}f_{\lambda\alpha} +
\frac{m^2}{2}(f^2 - f_{\mu\nu}f^{\nu\mu}) \rbrack
\label{3.1}
\ee
where $f = \eta^{\mu\nu}f_{\mu\nu}$. The metric is flat :$\eta^{\mu\nu}$= diag (-,+,+). 
In this model we use second rank tensor fields, like $ f_{\alpha\beta}$ with no symmetry with respect to
their indices. Replacing $m$  by $-m$ in (\ref{3.1}) implies helicity change from $+2$ to $-2$. 
The first term in (\ref{3.1}) is reminiscent of a spin one topological Chern-Simons term which will be
called a Chern-Simons term of first order.
The second term in (\ref{3.1}) is the Fierz- Pauli(FP) mass term \cite{F} which is the spin two analogue 
of a spin one Proca mass term. Note that FP term breaks the local invariance of the Chern-Simons term.
The above first order self dual massive spin-2 field action can be easily found after writing topologically 
massive gravity in an intrinsically geometric form language and then linearizing it \cite{AK1}.

Let us then consider the following doublet of first order lagrangian densities in presence of 
source terms ,

\be
{\cal{L}}(f) =   \frac{m}{2}\eps^{\mu\nu\lambda}{f_\mu}\,^{\alpha}\pa_{\nu}f_{\lambda\alpha} +
\frac{m^2}{2}(f^2 - f_{\mu\nu}f^{\nu\mu}) -\frac{m}{2}f^{\mu\nu}J_{\nu\mu}
\label{3.2}
\ee
\be
{\cal{L}}(g) =  - \frac{m}{2}\eps^{\mu\nu\lambda}{g_\mu}\,^{\alpha}\pa_{\nu}g_{\lambda\alpha} +
\frac{m^2}{2}(g^2 - g_{\mu\nu}g^{\nu\mu}) + \frac{m}{2}g^{\mu\nu}J_{\nu\mu}
\label{3.3}
\ee
where $f^{\mu\nu}$ and $g^{\mu\nu}$ are distinct fields. Note that although the helicites are $\pm2$,
the mass term is identical for both lagrangians (\ref{3.2}) and (\ref{3.3}). The case of different masses
will be dealt in the next section. Now the equations of motion are given by,
\be
{\eps_{\mu}}\,^{\nu\lambda}\pa_{\nu}f_{\lambda\alpha}  + m\,(f\eta_{\mu\alpha} - f_{\alpha\mu})= \frac{1}{2}J_{\alpha\mu}
\label{3.4}
\ee
\be
{\eps_{\mu}}\,^{\nu\lambda}\pa_{\nu}g_{\lambda\alpha}  - m\,(g\eta_{\mu\alpha} - g_{\alpha\mu})= \frac{1}{2}J_{\alpha\mu}
\label{3.5}
\ee
Following our previous approach, let us introduce new fields $F$ and $G$ as,
\begin{eqnarray}
F_{\mu\alpha} = f_{\mu\alpha} + g_{\mu\alpha}; \quad\quad G_{\mu\alpha} = f_{\mu\alpha} - g_{\mu\alpha};\nonumber\\
F =\eta^{\mu\alpha}f_{\mu\alpha}= f+g;\quad\quad G=\eta^{\mu\alpha}G_{\mu\alpha}=f-g
\label{3.5a}
\end{eqnarray}
Now adding (\ref{3.4}) and (\ref{3.5}) and substituting old fields by new ones defined in
(\ref{3.5a}) leads to
\be 
{\eps_{\mu}}\,^{\nu\lambda}\pa_{\nu}F_{\lambda\alpha}  - m\,( G_{\alpha\mu} -\eta_{\alpha\mu}G) = J_{\alpha\mu}
\label{3.6}
\ee
Our motivation now is to express the above equation solely in terms of the G-field. To achive this
we abstract certain results from (\ref{3.4}).

$\bullet$ 
Contraction by $\eta^{\alpha\mu}$ of (\ref{3.4}) yields
\be
{\eps^{\alpha\nu\lambda}}\pa_{\nu}f_{\lambda\alpha} + 2mf =\frac{1}{2}J;\quad\quad\quad 
J=\eta^{\mu\alpha}J_{\mu\alpha}
\label{3.7}
\ee

$\bullet$
 Contraction by $\eps^{\mu\alpha\rho}$ of (\ref{3.4}) leads to 
\be
  -\pa_{\alpha}f^{\rho\alpha} + \pa^{\rho}f -m\,{\eps^{\mu\alpha\rho}}f_{\alpha\mu}=
\,\frac{1}{2}{\eps^{\mu\alpha\rho}}J_{\alpha\mu }
\label{3.8}
\ee

$\bullet$ 
Operating (\ref{3.4}) by $\pa^\mu$ on both sides gives
\be
 -m\,(\pa^{\mu}f_{\alpha\mu} - \pa_{\alpha}f) = \frac{1}{2} \pa^{\mu}J_{\alpha\mu}
\label{3.9}
\ee

Taking the difference of (\ref{3.4}) and (\ref{3.5}) and exploiting (\ref{3.5a}) leads to 
\be
{\eps_{\mu}}\,^{\nu\lambda}\pa_{\nu}G_{\lambda\alpha}  - m\,( F_{\alpha\mu}  -\eta_{\alpha\mu}F) = 0
\label{3.10}
\ee
Taking the trace yields the identity,
\be
 F \,=\,-\,\frac{1}{2m}{\eps^{\mu\nu\lambda}}\pa_{\nu}G_{\lambda\mu} 
\label{3.11}
\ee
Using (\ref{3.11}) in (\ref{3.10}) we obtain 
\be
 F_{\alpha\mu} \,=\, \frac{1}{m}\eps_{\mu}\,^{\nu\lambda}\pa_{\nu}G_{\lambda\alpha} \,-\,
\frac{1}{2m}\eta_{\alpha\mu}[\eps^{\rho\sigma\omega}\pa_{\sigma}G_{\omega\rho}]
\label{3.12}
\ee

Now substituting $F_{\alpha\mu}$ in (\ref{3.6}) gives,
\be
  \frac{1}{m}{\eps_{\mu}}\,^{\nu\lambda}\pa_{\nu} [{\eps_\alpha}\,^{\rho\sigma}\pa_{\rho}G_{\sigma\lambda}\,
-\,\frac{1}{2} \eta_{\lambda\alpha}\eps^{\rho\sigma\omega}\pa_{\sigma}G_{\omega\rho}]
\,-\,m[ G_{\alpha\mu}  -\eta_{\alpha\mu}G] = J_{\alpha\mu }
\label{3.13}
\ee
Combining (\ref{3.7}) and  (\ref{3.8}) we obtain,
\be
\eps^{\mu\alpha}\,_{\rho}f_{\alpha\mu} =\frac{1}{2m^2}\pa^{\mu}J_{\rho\mu }-
\frac{1}{2m}\eps^{\mu\alpha}\,_{\rho}J_{\alpha\mu}
\label{3.13a}
\ee
The corresponding equation for $g$ follows by replacing $m$ by $-m$, 
\be
\eps^{\mu\alpha}\,_{\rho}g_{\alpha\mu} =\frac{1}{2m^2}\pa^{\mu}J_{\rho\mu }+
\frac{1}{2m}\eps^{\mu\alpha}\,_{\rho}J_{\alpha\mu}
\label{3.13b}
\ee
Subtracting (\ref{3.13b}) from (\ref{3.13a}) yields
\be
\eps^{\mu\alpha\rho}G_{\alpha\mu}=-\frac{1}{m}\eps^{\mu\alpha\rho}J_{\alpha\mu}
\label{3.13c}
\ee
Therefore from (\ref{3.13c}) we can conclude,
\be
\eps^{\mu\alpha\rho} \pa_{\rho} G_{\alpha\mu}=-\frac{1}{m}\eps^{\mu\alpha\rho} \pa_{\rho}J_{\alpha\mu}
\label{3.14}
\ee
\be
G_{\alpha\mu} - G_{\mu\alpha} = -\frac{1}{m}(J_{\alpha\mu} - J_{\mu\alpha} )
\label{3.15}
\ee
Substituting (\ref{3.14}) in (\ref{3.13}) we get 

\be
{\eps_{\mu}}\,^{\nu\lambda}\pa_{\nu} [{\eps_\alpha}\,^{\rho\sigma}\pa_{\rho}G_{\sigma\lambda}]-
{m^2}[ G_{\alpha\mu}  -\eta_{\alpha\mu}G] = -\frac{1}{2m}{\eps_{\mu\nu\alpha}}\pa^{\nu}
 [\eps^{\rho\beta\omega}\pa_{\beta}J_{\omega\rho}] + m J_{\alpha\mu}
\label{3.16}
\ee
The symmetrised version of the above equation reads,
\begin{eqnarray}
{\eps_{\mu}}\,^{\nu\lambda}\pa_{\nu} [{\eps_\alpha}\,^{\rho\sigma}\pa_{\rho} G_{\sigma\lambda}]
+{\eps_{\alpha}}\,^{\nu\lambda}\pa_{\nu} [{\eps_\mu}\,^{\rho\sigma}\pa_{\rho}G_{\sigma\lambda}]
-{m^2}[ G_{\alpha\mu}+G_{\mu\alpha}] \cr
 + 2m^{2}\eta_{\alpha\mu}G= m (J_{\alpha\mu}+J_{\mu\alpha})
\label{3.16a}
\end{eqnarray}
Exploiting (\ref{3.15}) we obtain the final effective equation of motion,
\be
 \frac{1}{2}{\eps_{\mu}}\,^{\nu\lambda}\pa_{\nu} [{\eps_\alpha}\,^{\rho\sigma}\pa_{\rho}(G_{\sigma\lambda}+
G_{\lambda\sigma})] - {m^2}[ G_{\alpha\mu}  -\eta_{\alpha\mu}G] = m J_{\mu\alpha}
\label{3.17}
\ee
Let us now discuss, in the absence of sources, the factorisability of the equations of motion. Some
conditions on the tensor field are necessary to achieve this factorisation. It is known \cite{DM1}
from a study of the equations of motion of (\ref{3.1}) that the tensor field $f_{\mu\nu} $ satisfies
 $a)$ tracelessness $ f^{\mu }_{\mu}=0$, $b)$ transversality $ \pa^{\mu }f_{\mu\nu}=0$ and $c)$ symmetricity
$f_{\mu\nu} = f_{\nu\mu}$. Consequently the composite fields $F_{\mu\nu},G_{\mu\nu} $ in (\ref{3.5a})
should also satisfy these properties. Indeed one may also verify this directly from (\ref{3.17}), in the absence
of sources. Under these conditions (\ref{3.17}) factorises as,
\be
(-\eps^{\mu\rho\beta}\pa_{\rho} + m \eta^{\mu\beta})(\eps_{\mu}\,^{\nu\lambda}\pa_{\nu} - m \eta^{\lambda}_{\mu})
G_{\lambda\alpha}=0
\label{4b.3}
\ee
We observe that the above equation of motion (\ref{3.17}) corresponds to an effective theory whose action is given by
\be
S = \int{d^3}x\lbrack \frac{1}{4}G.d\Omega(G)   + \frac{m^2}{2}(G^2 - G_{\mu\nu}G^{\nu\mu})-\frac{1}{2}m
G_{\mu\nu}J^{\nu\mu}  \rbrack
\label{3.18}
\ee 
where 
$$ G.d\Omega(G) = G^{\mu\alpha}{\eps_{\mu}}\,^{\nu\lambda}\pa_{\nu} [{\eps_\alpha}\,^{\rho\sigma}\pa_{\rho}
(G_{\sigma\lambda}+G_{\lambda\sigma})]\nonumber
$$
\\
Note that the first term in the action (\ref{3.18}) stands for the quadratic Einstein-Hilbert term
while the second one is the Pauli-Fierz mass term applicable for spin 2 particle. In the absence
of source this action corresponds to an effective theory which is the analogue of Proca
model for spin-1 case in vector theory. 

Proceeding in a likewise manner the equation of motion for the $F$-field emerges as, 

\begin{eqnarray}
 \frac{1}{2}{\eps_{\mu}}\,^{\nu\lambda}\pa_{\nu} [{\eps_\alpha}\,^{\rho\sigma}\pa_{\rho}(F_{\sigma\lambda}+
F_{\lambda\sigma})] - {m^2}[ F_{\mu\alpha}  -\eta_{\alpha\mu}F] = 
m [\eps_{\alpha\mu\rho}\pa_{\omega} J^{\rho\omega} + \cr
\eps_{\mu\nu\lambda}\pa^{\nu} J^{\lambda}\,_{\alpha}+\eps_{\alpha\nu\lambda}\pa^{\nu} J^{\lambda}\,_{\mu}]
\label{3.20}
\end{eqnarray}

In the absence of sources this is just a replica of (\ref{3.17}). Thus, as happened for the vector
model, either combination $F$ or $G$ yields an effective theory which has the Einstein-Hilbert term
and the F-P term with differences cropping in the source terms.

The analogue of the map (\ref{2.10}) is now written for the spin $2$ example,

\be
{\cal L}_{2SD}(f,g) \Longleftrightarrow {\cal L}_{EHFP}(f \pm g)
\label{3.20a}
\ee
Here the doublet of self dual models on the left hand side is given by (\ref{3.2},\ref{3.3}) while the composite
Einstein-Hilbert Pauli-Fierz model is defined in (\ref{3.18}).
\section*{V. Tensor fields with distinct mass}

In this section we repeat the analysis for the doublet (\ref{3.2}) and(\ref{3.3})
but with distinct mass parameters. To avoid technical complications we drop the source terms. 
We show that combining this doublet yields an effective theory that has an E-H term, a FP mass term
and a generalised first order CS term. This CS term contains, apart from the standard form
given in  (\ref{3.1}), two other similar terms with a different orientation of indices.
Consider therefore the lagrangian densities, 

\be
{\cal{L_+}}(f) =   \frac{m_1}{2}\eps^{\mu\nu\lambda}{f_\mu}\,^{\alpha}\pa_{\nu}f_{\lambda\alpha} +
\frac{{m_1}^2}{2}(f^2 - f_{\mu\nu}f^{\nu\mu}) 
\label{3.21}
\ee
\be
{\cal{L_-}}(g) =  - \frac{m_2}{2}\eps^{\mu\nu\lambda}{g_\mu}\,^{\alpha}\pa_{\nu}g_{\lambda\alpha} +
\frac{{m_2}^2}{2}(g^2 - g_{\mu\nu}g^{\nu\mu}) 
\label{3.22}
\ee
Now (\ref{3.21}) and (\ref{3.22}) yield the equations of motion,
\be
{\eps_{\mu}}\,^{\nu\lambda}\pa_{\nu}f_{\lambda\alpha}  + {m_1}\,(f\eta_{\mu\alpha} - f_{\alpha\mu})= 0
\label{3.23}
\ee
\be
{\eps_{\mu}}\,^{\nu\lambda}\pa_{\nu}g_{\lambda\alpha}  - {m_2}\,(g\eta_{\mu\alpha} - g_{\alpha\mu})= 0
\label{3.24}
\ee
Following identical field definitions as (\ref{3.5a}) and analogous steps, 
it can be shown that the final form of the equation of motion for $ G_{\mu\nu} $ is given by,

\begin{eqnarray}
- \frac{1}{2}{\eps_{\mu}}\,^{\nu\lambda} \pa_{\nu} [{\eps_\alpha}\,^{\rho\sigma}\pa_{\rho}(G_{\sigma\lambda}+
G_{\lambda\sigma})] - {m_1}{m_2}[ G_{\alpha\mu}  -\eta_{\alpha\mu}G]  + \cr 
\frac{1}{2}({m_1}-{m_2})({\eps_{\mu}}\,^{\nu\lambda}\pa_{\nu}G_{\lambda\alpha }  
 + {\eps_{\alpha}}\,^{\nu\lambda}\pa_{\nu}G_{\lambda\mu}) =0
\label{3.25}
\end{eqnarray}
A similar equation of motion is also obtained for the other variable $F_{\mu\nu} $. 

The action from which the above equation of motion (\ref{3.25}) follows is given by,

\begin{eqnarray}
S = \int{d^3}x\lbrack \frac{{m_1}{m_2}}{2}(G^2 - G_{\mu\nu}G^{\nu\mu})+
\frac{1}{4}G.d\Omega(G)   
+\frac{1}{8}({m_2}-{m_1}) \cr
\{{\eps_{\mu}}\,^{\nu\lambda}G^{\mu\alpha}\pa_{\nu}G_{\lambda\alpha } 
+ {\eps_{\mu}}\,^{\nu\lambda}G^{\alpha\mu}\pa_{\nu}G_{\alpha\lambda}  
+ \,2 {\eps_{\mu}}\,^{\nu\lambda}G^{\alpha\mu}\pa_{\nu}G_{\lambda\alpha } \} ]
\label{3.26}
\end{eqnarray}

We have thus successfully combined different mass terms in the spin 2 case to yield
the action (\ref{3.26}) of the effective theory.
While the first two terms are the usual FP and E-H terms the last piece, which is a consequence
of different masses, is a generalised form of the CS term. As announced earlier it has, apart
from the usual structure, two other pieces that may be obtained from a reorientation of indices. 
In fact it has all possible orientations of indices leading to a first order Chern-Simons term.
Furthermore if we impose a condition of symmetricity $G^{\mu\alpha}=G^{\alpha\mu} $, then all
pieces become identical and the standard first order C-S term with a coefficient $\frac{1}{2}({m_2}-{m_1})$
is obtained.
The first term in (\ref{3.26}) is the Fierz-Pauli(FP) mass term with mass co-efficient $m=\sqrt{{m_1}{m_2}}$.
The second term involves the usual kinetic term (defined in the previous section) which is
equivalent to linearized Einstein-Hilbert(EH) term upto quadratic order.
Thus the action (\ref{3.26}) for spin-2 particle may be interpreted as an analogue 
of Maxwell-CS-Proca model for spin-1 particle. Incidentally the C-S term for the vector case
has a unique orientation of indices $\eps_{\mu \nu\lambda}f^{\mu}\pa ^{\nu} f ^\lambda$ and any
changes are absorbed in a trivial normalisation of signs.

\section*{VI. Conclusion}

The present analysis depicts the important role of symmetry in understanding various models
in odd dimensions. The dual nature of symmetry manifested in (left-right) chirality
or (anti)self duality was responsible for the properties of the final theory. For example,
a two dimensional oscillator could be interpreted as a composition of two chiral oscillators
moving in opposite directions. Chirality therefore gets hidden in an ordinary two dimensional
oscillator since the opposing effects of chirality in its constituent pieces are cancelled.
Indeed the generalised Landau problem with electric and magnetic fields was shown to be composed
of such chiral oscillators. The explicit demonstration was done at the level of equations of
motion with an appropriate change of variables.

The quantum mechanical example served as the bedrock from where the more involved examples of field theory
and gravity were studied. More specifically, the similarity in the structures of the 
quantum mechanical model and the other models in field theory/gravity naturally suggested
this possibility of dual composition. Once this was evident the rest was more a matter
of technique. Self and anti-self dual models in $(2+1)$ dimensions combined to yield
the Maxwell-Chern-Simons-Proca model. In the case of gravity we obtained a new form
of generalised self dual model. The correct sign of the Einstein-Hilbert term was obtained. Apart from this there were
two mass terms. One was the usual Fierz-Pauli term while the other was a generalised form of the first order
Chern-Simons term that encompassed all possible permutations of the indices. If we imposed the condition
that the rank two tensor field was symmetric then the self dual model discussed in \cite{DM} was reproduced.

We have also discussed the factorisability of equations of motion of different models. Such a phenomenon
illuminates the dual composition of the models. Specially in case of gravity, this factorisation is 
possible subject to certain conditions following from the equation of motion.

In hindsight it might be desirable, though not essential, to visualise in general terms the obtention
of a new theory from a combination of chiral ones. Chiral theories occur in doublets corresponding to 
the left and right degrees of freedom. The equations of motion following from this doublet are form invariant,
differing only by a sign in the chiral piece. Adding and subtracting these equations naturally leads to
a combination which is either a sum or a difference of the original variables. Renaming this 'sum' and 'difference'
as new fields yields a pair of coupled differential equations. It is then possible to eliminate one of these
new fields in favour of the other using these differential equations. The final outcome is an equation
of motion involving only the new fields. Furthermore, the symmetrical treatment implies that we obtain
identical equations of motion for both the new fields. Consequently we are led to a unique new theory
obtained by a composition of the chiral degrees of freedom. 
 
We feel that our approach is simple and economical when compared with other approaches \cite{DM,DM1,BH}
which discuss such a dual structure in odd dimensional theories. Contrasted with existing lagrangian or
hamiltonian based approaches that require involved field redefinitions or canonical transformations, 
the formulation here is tranparent as well as generic. The simple change of variables
(\ref{2c}) is universally applicable and irons out obstacles faced otherwise in treating field 
theoretical or gravity models. Also, contact or interference terms were completely avoided
to manifest the dual structure. Inclusion of sources posed on problems. We feel that the present
illustration could be useful in unravelling other features of gravitational models. 
\vskip 0.5cm

{\bf{Acknowledgement}}\\
One of the authors Sarmishtha Kumar(Chaudhuri) is grateful to S.N.Bose National Centre For Basic Sciences  
for providing necessary facilities.


\end{document}